# Structure Learning in Coupled Dynamical Systems and Dynamic Causal Modelling

*Amirhossein Jafarian\*, Peter Zeidman, Vladimir Litvak and Karl Friston*

*The Wellcome Centre for Human Neuroimaging, Institute of Neurology, 12 Queen Square, London, UK. WC1N 3AR*

## Summary

Identifying a coupled dynamical system out of many plausible candidates, each of which could serve as the underlying generator of some observed measurements, is a profoundly ill-posed problem that commonly arises when modelling real world phenomena. In this review, we detail a set of statistical procedures for inferring the structure of nonlinear coupled dynamical systems (structure learning), which has proved useful in neuroscience research. A key focus here is the comparison of competing models of network architectures – and implicit coupling functions – in terms of their Bayesian model evidence. These methods are collectively referred to as dynamic casual modelling (DCM). We focus on a relatively new approach that is proving remarkably useful; namely, Bayesian model reduction (BMR), which enables rapid evaluation and comparison of models that differ in their network architecture. We illustrate the usefulness of these techniques through modelling neurovascular coupling (cellular pathways linking neuronal and vascular systems), whose function is an active focus of research in neurobiology and the imaging of coupled neuronal systems.

## Introduction

This paper sets out a general method for addressing the problem of *structure learning*; namely, identifying a coupled dynamical system that best accounts for empirical observations (1,2). In this context, a hypothesis about the structure of a system, for example the connectivity architecture of a neural network, is expressed formally as a *model* (3). The objective is to search over models (e.g., by pruning redundant connections) to arrive at a network architecture that optimally explains the data. This approach rests upon dynamic casual modelling (DCM) (4), Bayesian model selection (BMS) (5) and Bayesian model reduction (BMR) (6), which are implemented in the freely available software called Statistical Parametric Mapping (SPM) (7) (see Table 1 for further description of terminology used in this paper). These methods have been developed for the analysis of large scale recordings of brain activity, however they could be conveniently applied in other domains, where mechanistic modelling based on empirical data is of interest.

The basic idea behind DCM is to convert data assimilation, identification and structure learning problems into a generic inference problem – and then use variational Bayesian techniques to infer all the unknowns; ranging from unobservable or latent states through to the structure or form of the dynamical system that best accounts for the data at hand. In other words, DCM enables qualitative and quantitative questions to be asked about the dynamical system generating data (usually timeseries).

\*Author for correspondence (a.jafarian@ucl.ac.uk).   ORCID  ID: 0000-0001-9547-7941.

†Present address: The Wellcome Centre for Human Neuroimaging, 12 Queen Square, University College London, WC1N 3AR, UK.

Formally, DCM is the Bayesian inversion of a biophysically informed dynamical (i.e., state space) model, given some timeseries data; usually, neuroimaging data (M/EEG, fMRI). The models that underwrite DCM are generally specified in terms of (ordinary, stochastic or delay) differential equations describing the coupling within and among dynamical systems. Hypotheses about the architecture of (directed) coupling are tested by comparing the evidence for the data under different models (4). DCM can be employed to infer coupling (within each node and between nodes) of a nonlinear dynamical system, given empirical responses of a system (i.e., the brain) to experimental inputs or different dynamical regimes. This proceeds by inferring condition-specific parameters, which explain/model how changes in experimental context (or different dynamical regimes) are mediated by changes in particular connections in the model. This ability to model context sensitive changes in coupling distinguishes DCM from other *data assimilation* and *identification* methods (e.g., *parameter tracking* (8), *compressive sensing based dynamical system identification* (9) *and time-evolving dynamical Bayesian inference of coupled dynamical systems* (10), *to name a few*). The outcomes of model inversion using DCM are a posterior probability density over model parameters (that parameterise coupling and context sensitive effects) and the relative probability of having observed the data under each model (model evidence). This model evidence or marginal likelihood can be used to draw conclusions by comparing the evidence for different models – known as Bayesian model comparison and selection (BMS).

In the past two decades, DCM for functional magnetic resonance imaging (fMRI) has been applied in many studies in the field of cognitive neuroscience (e.g., aging (11), memory (12)) as well as psychiatric disorders (13–15). In parallel, using the same conceptual principles, DCM has also been applied to Magneto/Electroencephalography (M/EEG) data, to disambiguate neuronal causes of electromagnetic responses such as induced responses (16), phase coupling (17), event related potentials (18,19) and also to provide insights into underlying generators of neurological disorders (20–22). More recently, DCM has motivated and contributed to the development of research in theoretical neuroscience such as predictive coding (23), active inference (24) and, interestingly, the Bayesian brain hypothesis, which aims to establish the mathematical foundations of how the brain interacts with – and understands – its environment (25).

In this paper, we review and illustrate recent developments in Bayesian inference that enable an efficient procedure for learning the structure of coupled dynamical systems. First, we present the theoretical foundations of structure learning – using DCM – in a general form that may have wider application in engineering, physics and mathematical biology. We then present an example that highlights the usefulness of structure learning in the field of neuroscience. All software developments relating to the results in this paper are freely available through the academic SPM software (https://www.fil.ion.ucl.ac.uk/spm/). A glossary of technical terms used in this paper is provided in Table 1.

# Dynamic Causal Modelling and structure learning

The pipeline for studying the underlying generators of neuroimaging data using DCM is shown in Figure 1. The procedure begins by designing and conducting an experiment to study some particular function of the brain. Data features are then selected from the measured data and one or more hypotheses are formally expressed as (biophysically informed) coupled dynamical systems. A Bayesian (variational Laplace) scheme is then used to infer the settings of the parameters for each model (e.g. coupling strengths) and to quantify each model's evidence. Structure learning is then performed to compare the alternative model architectures, using BMS and BMR; in order to identify the best explanation for the underlying generators of the data. In the following sections we describe each of these steps in turn, before turning to a worked example.

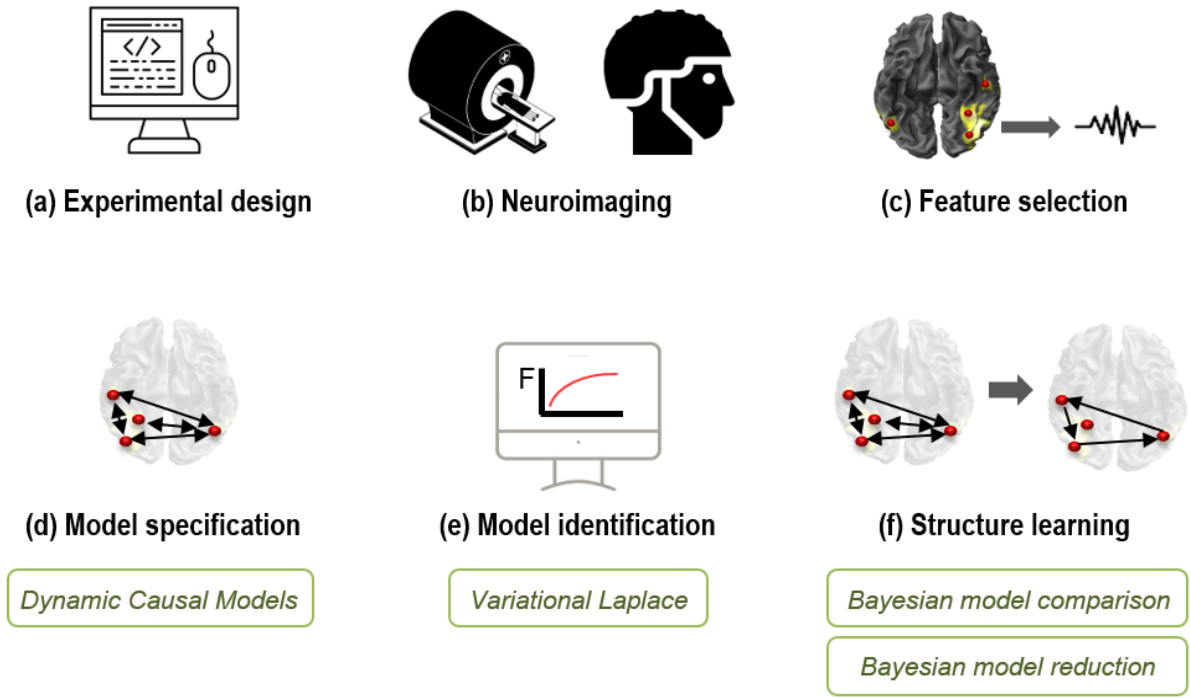

*Figure 1. **Structure learning using dynamic causal modelling.** a) The pipeline begins by designing an experiment to study a functional aspect of the brain, b) followed by recording brain activity using neuroimaging devices, such as MEG or fMRI. c) Feature selection is performed on the neuroimaging data; for example, by calculating evoked responses by averaging over trials, or transforming neuroimaging data to the frequency domain. d) Having prepared the data, the experimenter postulates several hypotheses (specified as models) about the underlying generation of the neuroimaging data. These can be expressed in terms of connections between and within brain regions (effectivity connectivity). e) These models are fitted to the data by finding the setting of the parameters that optimise variational free energy (F). f) The evidence associated with each model is compared using Bayesian model selection and/or reduction, to identify the most likely structure that accounts for the data. Image credits: MRI scanner by Grant Fisher, TN, screen in (a) and (e) by Dinosoft Labs, PK, all from the Noun Project, CC BY 3.0.*

## Experimental design

A DCM study begins by carefully articulating hypotheses and designing an experiment to test them. Typically, to maximise efficiency, there will be two or more independent experimental manipulations at the within-subject level, forming a factorial design. There may be additional experimental factors at the between-subject level, for instance to investigate differences between patients and controls, which will inform the strategy for sampling participants from the wider population. The hypotheses determine the choice of neuroimaging modality (e.g., M/EEG, fMRI), as well as the data features that will be selected and the type of dynamic causal model that will be used.

## Feature selection

The next step is to select features in the collected data that are important (i.e., informative) from a modelling standpoint. This is known as feature selection or extraction. For example, averaging time series over trials in response to stimulation gives event related potentials (ERPs) (26), or neuroimaging data can be represented in the frequency (27,28) or time-frequency domain (16).

## Model specification

The hypotheses are then formally expressed in terms of $k$ distinct and biologically informed (state space) model architectures $M = \{m_1, ..., m_k\}$, each describing possible interactions between experimental inputs and neuronal dynamics. In effect, a model in DCM can be understood as a dynamical system distributed on a graph, where the neuroimaging data captures the activity of each node (either directly – as in fMRI or indirectly via some mapping to sensor space – as in EEG). Depending on the scale and fidelity of the neuroimaging measurement, each node could, in principle, correspond to a compartment of a neuron, or to an individual neuron, or (more typically) to the average activity of millions of neurons constituting a *neuronal population* or a *brain region*. At any given scale, connections between nodes in this graph are referred to as the *effective connectivity;* namely, the effect of one node on another.

In general, partially observed discrete-continuous dynamical systems (which commonly arise in many mathematical and engineering applications) are well suited for modelling neural interactions (29). The generative (state space) model in DCM can be written as follows (30):

$$\begin{aligned} \dot{z} &= f(z, U, \theta^{(f)}) \\ Y &= g(z, \theta^{(g)}) + c(\beta_0) + \epsilon \end{aligned} \quad (1)$$

The first line in equation 1 governs dynamics of interactions (coupling) within and between nodes of the coupled dynamical system, where $z$ are (usually unobservable or hidden biological) states with a flow that depends upon parameters $\theta^{(f)}$ and exogenous or experimental input $U$. When the exogenous inputs are random fluctuations or innovations, equation 1 becomes a stochastic differential equation. The choice of coupling function $f$ (parametrised in terms of extrinsic and intrinsic coupling) is usually motivated by biological principles e.g., (31,32). The second line in equation 1 is known as the observer function, and links (usually observable neuroimaging) data $Y$ to the hidden or latent variables e.g., (33,34). In the second line of equation 1, the function $g$ models the contribution of the hidden states (depending upon parameters $\theta^{(g)}$) to the data. The second term in the observer equation, $c(\beta_0)$, models confounding signal components (e.g., drift) where $c(.)$ is typically a general linear model with parameters $\beta_0$ (e.g., mean of signal) (30,35). In the observer model, $\epsilon$, denotes the measurement error, which is conventionally modelled by a zero mean identically independent process (I.I.D), the covariance of which is estimated from the data. Hereinafter, unknown parameters in equation 1 are denoted by $\theta$ which includes model parameters $\theta^{(f)}$, observation function parameters, $\theta^{(g)}$, and parameters $\lambda$ that model the covariance of the observation noise (detailed in the supplementary material). The dynamics of each node are governed by (a set of) differential equations and the nodes are connected to each other. In effect, DCM estimates the parameters associated with the dynamics (i.e., differential equations) of each node and the coupling between them. There are many forms of models that have been used in DCM, for example, biophysical hemodynamic models (4,33,36–38), neuronal mass (19,31,39) and field models (40), weakly coupled oscillators (17), etc.

## Model identification

Given a prior probability density over the unknown parameters, $p(\theta)$, initial states, $z(0)$, and neuroimaging data $Y$, DCM is used to infer the posterior probability of the parameters of each model, using a gradient descent on variational free energy, $F$, as its cost function (please see supplementary material for further information). The free energy, also known as the evidence lower bound in machine learning, is a lower bound on the log *model evidence* (or marginal likelihood) $p(Y|m)$ (*m* denotes the model). In general, log model evidence is an unknown value, which can be decomposed as follows (30,41):

$$\ln p(Y|m) = F + D_{kl}(q(\theta), p(\theta|Y)) \qquad (2)$$

In equation 2, $D_{kl}(q(\theta), p(\theta|Y))$ is the Kullback-Leibler (KL) divergence between the approximate and true posterior over the parameters, which are denoted by $q(\theta)$ and $p(\theta|Y)$, respectively. F is (variational) free energy, defined as the difference between the accuracy of the model (i.e., expected log likelihood of the model, $E_{q(\theta)}[\ln p(Y|m)]$) and the complexity of the model (i.e., KL divergence between the prior, $p(\theta|m)$ and approximated posterior, $q(\theta)$, over the parameters, $D_{kl}(p(\theta|m), q(\theta))$) (30,42):

$$F \cong \mathrm{E}_{q(\theta)}[\ln p(Y|m)] - D_{kl}(q(\theta), p(\theta|m)) \qquad (3)$$

Given that the log-model evidence - the left hand side of equation 2 - is fixed (although unknown), by maximizing the free energy we minimize the divergence between the approximate and true posterior. The free energy $F$ scores the goodness of a hypothesis (i.e., model), which can be employed for model comparison and for inferring model parameters – where the approximate posterior density $q(\theta)$ quantifies beliefs about the parameters. To identify the setting of the parameters that maximises free energy, DCM uses an estimation scheme known as variational Laplace; namely, variational Bayes under the Laplace approximation (i.e., the prior and posterior densities of latent variables have Gaussian distributions). Intuitively, maximising the free energy offers a form of regularisation, because the KL divergence term (complexity) in equation 3 acts as a penalty. Therefore, as the number of parameters increases – or the (approximate) posterior over parameters deviates from the prior – the free energy decreases. This finesses the risk of overfitting or implausible parameter estimates. The KL term in equation 3 accounts for both the expected values of the parameters and their covariance, providing a closer approximation of the log evidence than the Akaike information criterion (AIC) or Bayesian information criterion (BIC), which define complexity as (effectively) the number of parameters, without considering their posterior covariance (43).

To summarise, model identification in DCM is carried out by gradient descent of a variational free energy functional under the Laplace approximation (i.e., both prior and posterior densities over parameters have a Gaussian form) to identify the parameters of dynamical systems distributed on a graph. Usually, the form of the (ordinary, partial, stochastic or delay) differential equations and priors over their parameters are chosen to embody biophysical constraints. These constraints usually guarantee a degree of dynamical stability, which is important during model inversion; particularly when using approximate Bayesian inference. This is because brittle or highly nonlinear systems violate the Laplace assumption inherent in the inversion schemes commonly used in DCM. Generally, the dynamics on each node have simple point attractors

(associated with low order Taylor expansions of nonlinear differential equations) (44). Having said this, it is possible to consider autonomous dynamics on each node using phase reduction techniques. For example, DCM for phase coupling uses hidden neuronal states that are the phase of weakly coupled oscillators, with implicit (quasi) periodic dynamics (17,45).

The scheme described above provides a systematic way of handling the many scenarios where model parameters change over time. The simplest example would be the condition specific effects modelled by bilinear coupled dynamical systems (37) which were first introduced in the context DCM for fMRI (4). With time varying inputs, bilinear effects entail that the coupling (*effective connectivity*) is itself a function of time and therefore context sensitive. Another important example arises in the case of modelling epileptic seizures; where a slow drift of physiological parameters might give rise to changes of brain states, from apparently normal to pathological conditions (e.g., via phase transitions and bifurcations) (22). In this case, hierarchical linear regression of the parameters, estimated over successive windows of data, can be employed to identify coupling that varies with specified experimental variables (46). The first level of the hierarchical model corresponds to the coupling estimated within each time window. The second level of the model uses the ensuing coupling (from the first level) to model changes (here fluctuations) of the parameters across windows. This approach has proven useful in the analysis of longitudinal resting state fMRI data in human neurosurgery patients (47) and for modelling seizure activity in rodent (48) and zebrafish (22) models.

These model identification methods facilitate testing hypotheses in an experimental setting. They also enable the validation of new models using the DCM framework. Conventionally, models in DCM are validated in three ways; namely, (i) face validity, where simulated data is used to ensure known model parameters (and structures) can be recovered following model inversion (ii) construct validity, where the inferences from DCM are compared with other (comparable) approaches, (iii) predictive validity, which tests whether the posterior predictions of DCM reflect a known or expected effect e.g. a pharmacological effect or diagnostic group membership.

## Structure learning

Having inverted a model to obtain its evidence and parameters, the next step is to ask whether the structure of the model could be simplified to further optimise the variational free energy. Recall that the free energy quantifies the trade-off between the accuracy and complexity of the model – so if a change to the network structure increases free energy, then the model has become more accurate (better fitting the data) and / or less complex (simpler in terms of its parameterisation). This process of selecting between network architectures (a.k.a. structure learning) depends on BMS; namely, the selection among different models based on their evidence. This process can be performed automatically and rapidly over potentially thousands of alternative models, using an approach called Bayesian model reduction.

The hypotheses or models $m_1, m_2, \ldots, m_k$ with free energy $F_1, F_2, \ldots, F_k$, can be compared using Bayesian model comparison. For any two inverted models, $m_i$ and $m_j$ with free energies of $F_i$ and $F_j$ respectively, the log Bayes factor, $\log B_{ij}$ is defined as follows (5,49):

$$\begin{aligned} \ln B_{ij} &= \ln p(Y|m_i) - \ln p(Y|m_j) \\ &= (F_i - D_{kl}[q(\theta), p(\theta|Y)]_i) - (F_j - D_{kl}[q(\theta), p(\theta|Y)]_j) \\ &\cong F_i - F_j \end{aligned} \qquad (4)$$

By estimating the parameters of each model (thereby maximising the free energy), the Kullback-Leibler divergence between the approximate and true posterior of the parameters vanishes in the limit (42,43). Therefore, the log Bayes factor can be approximated as the difference between free energies of the models. Conventionally, a log Bayes factor above three indicates that there is 'strong evidence' for model $m_i$ over model $m_j$ (5,49). Bayesian model comparison thereby allows pairwise model comparison, which in turn can be used to identify which of two models best accounts for the data. The posterior probability for each model can then be computed by application of Bayes rule. Under equal priors for each model, this simplifies to a logistic (sigmoid) function of the log Bayes factor:

$$p(m_i|Y) = \frac{1}{1 + \exp(-\ln BF_{ij})} \qquad (5)$$

This process is easily generalized to comparisons with more than two models, by computing the log Bayes factor of each model relative to any one of the models in the comparison set. In some studies, rather than one model corresponding to one hypothesis, it can make more sense for a set or *family* of models to represent a particular hypothesis. This generalisation of BMS – where the model space is grouped in several classes or families – is referred to as family-wise model comparison (5). To explain this, let model space $M = \{m_1, m_2, \dots m_k\}$ be grouped into $r$, disjoint model spaces $\{M_i\}_{i=1}^r \in 2^M$ where $\sum_{k=1}^r |M_i| \leq |M|$. In this case, the prior probability of each class is $\frac{1}{r}$. Consequently, the prior probability that a model $j$ belongs to the class $M_i$ (with cardinal $l$) is $P(m_j) = \frac{1}{rl}$. By applying Bayes' rule over the space of all models, one can calculate the posterior probability of model $m_j$ with evidence $p(Y|m_j)$, as follows:

$$p(m_j|Y) = \frac{p(Y|m_j)p(m_j)}{\sum_{k=1}^r p(Y|m_k)p(m_k)} \qquad (6)$$

By definition, the posterior probability for each family (i.e., class) is the sum of the posterior probabilities of its constituent models. The ensuing family posterior probabilities can then be compared using Bayesian model comparison. One ubiquitous application of family comparison is to compare models with and without a common feature or property – for example, with or without a particular parameter. In this case, BMS can be performed on the model space comprising two groups of models, where the properties of interest appear in only one group. This can be used to establish whether such a property is necessary to explain the data at hand.

By using the statistics detailed above, one can assign a probability to each of a pre-defined set of hypotheses about the structure of the neural network and draw conclusions based upon a small number of plausible architectures. An alternative approach to structure learning is to apply Bayesian model comparison in what can be called a *discovery mode*. By automatically searching over potentially thousands of alternative model architectures, one can ask whether eliminating any substructure (i.e., subset of parameters) of the model would increase the free energy relative to the original (full) model. This is made possible by a recent development known as Bayesian model reduction (BMR) (6,50). Assume a (full) model $m$ is fitted to data $Y$ with prior $p(\theta|m)$ and posterior $p(\theta|Y,m)$ with the parameter vector $\theta$. Using Bayes' rule (we have dropped the dependency on the model $m$ for clarity):

$$p(\theta|Y) = \frac{p(Y|\theta)p(\theta)}{p(Y)} \tag{7}$$

Where the model evidence is $p(Y) = \int p(Y|\theta)p(\theta)$, the log of which is approximated by the variational free energy. Crucially, the priors determine which parameters (e.g., connections) should be informed by the data. Having estimated the parameters and free energy of the model, which we will refer to as the 'full' model, BMR provides an analytic and rapid technique for evaluating the relative evidence for an alternative model, which differs only in terms of the priors $p(\tilde{\theta})$. Typically, this alternative set of priors will fix certain parameters to their prior mean, thereby reducing or pruning the model structure. For this reduced model, the approximate posterior $p(\tilde{\theta}|Y)$ of the parameters under the reduced priors is again given by Bayes rule:

$$p(\tilde{\theta}|Y) = \frac{p(Y|\theta)p(\tilde{\theta})}{\tilde{p}(Y)} \tag{8}$$

The likelihood function $p(Y|\theta)$ for the reduced and full models is the same, which enables equations 7 and 8 to be linked as follows:

$$p(\tilde{\theta}|Y) = p(\theta|Y)\frac{p(Y)\,p(\tilde{\theta})}{\tilde{p}(Y)\,p(\theta)} \tag{9}$$

Next, to find the evidence of the reduced model, both sides of equation 9 are integrated over the parameter space. Using the fact that $\int p(\tilde{\theta}|Y)\,d\theta = 1$, the model evidence for the reduced model is as follows:

$$\tilde{p}(Y) = p(Y)\int p(\theta|Y)\frac{p(\tilde{\theta})}{p(\theta)}\,d\theta \tag{10}$$

Equations 9 and 10 have analytic solutions given Gaussian forms for prior and posterior densities, meaning that the coupling parameters and evidence for reduced models can be derived from those of the full model in milliseconds, on a typical desktop computer (51). This speed is leveraged in DCM for automatically discovering an optimal coupling structure for a network. In this setting, a greedy search is used, which iteratively generates candidate reduced models with different priors (52). Their evidence is evaluated using BMR, and Bayesian model comparison is used to assess whether they should be retained or discarded. The ensuing coupling parameters from the best candidate models are averaged (using Bayesian model averaging based upon the evidence for each model) and returned as the optimal network structure for those data.

It is worth mentioning that, in our experience, a greedy search over reduced models performs well, and gives similar results to manually defined sets of reduced models (52). Nevertheless, other search/optimisation algorithms could be considered and their performance compared. The use of a greedy search is purely for computational expediency; in situations where the number of combinations of different parameters – that constitute distinct models – becomes prohibitive. When there are a reasonable number of models (e.g., in the hundreds), the model comparison (i.e., or reduction) can use an exhaustive search.

Bayesian model comparison may be contrasted against another method in the modelling literature – surrogate data testing, where the statistics of different data are compared (for instance, the statistics of empirical timeseries are compared against a null set of timeseries, generated by a process that lacks a particular parameter) (53). However, Bayesian model comparison operates at the level of models rather than summary statistics of the data, to provide a straightforward and efficient method for comparing models (without the need for sampling, if variational methods are used). Bayesian model comparison properly accommodates both model accuracy and complexity, meaning that any conclusions one draws will adhere to Occam's principle, i.e. the simplest explanation for the data should be favoured.

## Drawing conclusions

Together, the procedures described above constitute all the necessary tools for learning the optimal structure of a coupled dynamical system – as evinced by some data. We started by defining the free energy, a scalar functional that quantifies the trade-off between the accuracy and complexity of a model. This quantity is estimated in the context of a 'full' model with all coupling parameters of interest. Then, using BMR, the free energy for reduced models are derived analytically – either based on a few specified models – or by performing an automatic greedy search over potentially thousands of reduced models. Bayesian model comparison is used in both cases, to evaluate which model(s) should be favoured as explanations for the underlying generators of the data.

This concludes our overview of the basic approach to coupled dynamical systems. In what follows, we provide two worked examples to illustrate the sorts of inference and structure learning that is afforded. Although this example reflects our interest (i.e., physiological coupling in the brain) the analyses can, in principle, be applied to any coupled dynamical system that can be articulated in terms of (stochastic, ordinary or delay) differential equations.

# Worked Examples

To illustrate the theory reviewed in this paper, we consider modelling the neurovascular coupling system, which ensures that brain cells (neurons) are adequately perfused with oxygenated blood in an activity-dependent fashion. This illustrates dynamical coupling on two levels; first, the coupling between two different physiological systems (neuronal and haemodynamic systems). Second, the coupling among remote neuronal populations that underwrites distributed neuronal processing and representations. Understanding the functional architecture of neurovascular coupling facilitates our understanding of ischaemic brain injury (i.e., stroke) and, in research, establishes the origin of fMRI time series used in brain mapping. In short, neurovascular coupling determines how blood dynamics are altered due to neuronal demands for energy (35,54). The challenge in this field is that it is not possible to measure the activity of this system noninvasively in the human brain, and therefore modelling-based approaches have been widely applied.

In the following example, a coupled (biophysically informed) dynamical system that models the behaviour of neuronal responses and the vascular system was employed. This is illustrated for a single brain region (one node of the coupled system) illustrated in Figure 2. In this model, a canonical microcircuit (CMC) accounts for laminar-specific connections in a small area of cortex known as a cortical column. Dynamic causal models at this level of (neuronal) detail usually requires fast electrophysiological measurements such as M/EEG (17). Blood vessel dynamics caused by fast neuronal activity are captured by a haemodynamic model, the parameters of which are generally inferred using functional MRI (33). The intermediate system linking (fast) neuronal and (slow) haemodynamics is known as neurovascular coupling, which is the focus of the modelling work described below.

It is worth emphasising that the kind of model inversion problem illustrated here is extremely challenging, which is why it has driven multiple technical and theoretical developments. The single region model presented here comprises 12 unobserved biological states that operate on different time scales (the time resolution of the neuronal system and vascular dynamics are milliseconds and seconds respectively) with unknown coupling, as well as unknown parameters that link neuronal and vascular parameters to the observed fMRI and EEG recordings. The Bayesian methods reviewed here provide a useful basis for addressing profoundly ill posed problems in structure learning of coupled dynamical systems.

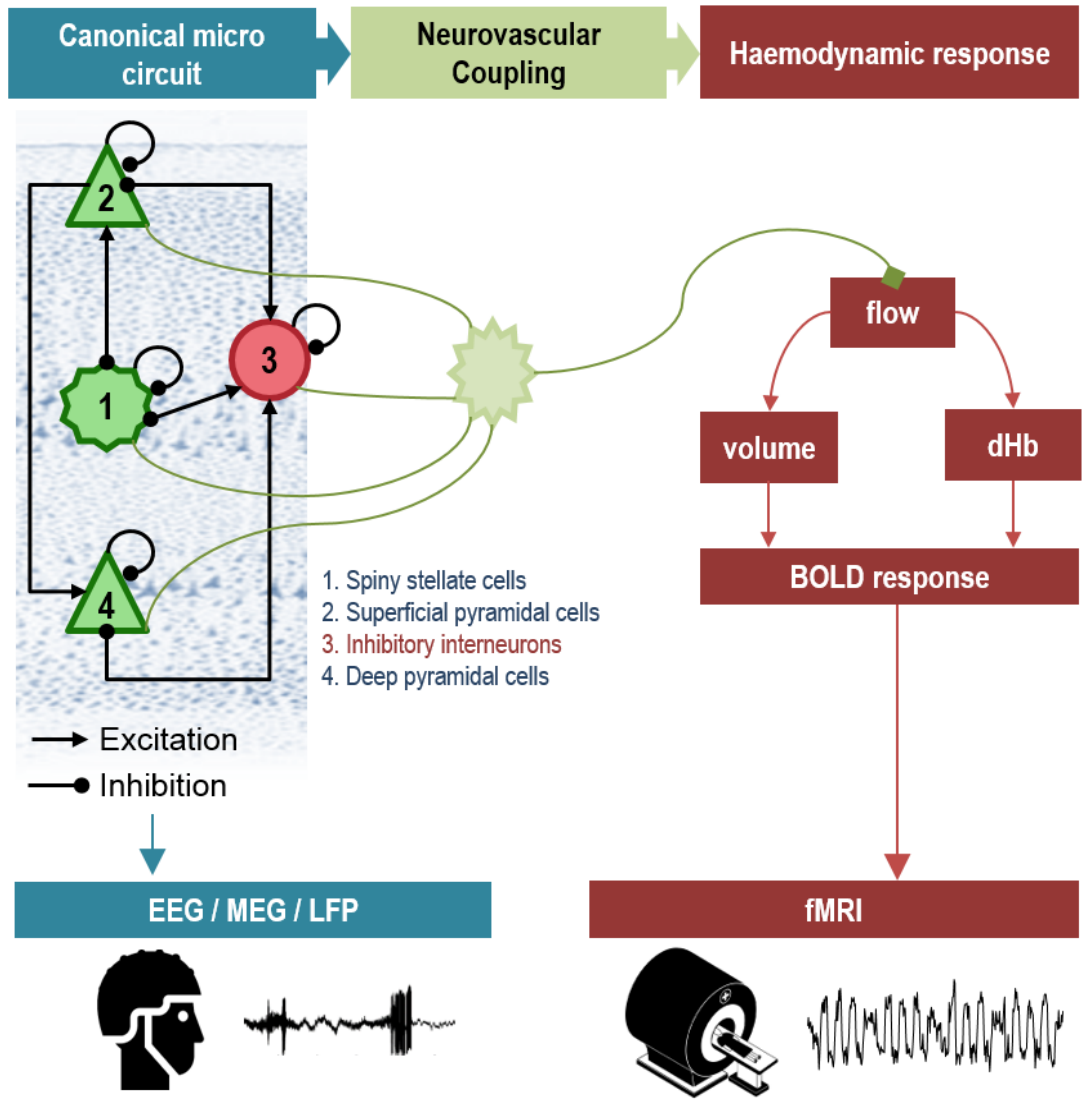

*Figure 2.* ***Macroscale model of neuronal to vascular pathway dynamics in a single brain region.*** *This model (34) couples neuronal electrical activity (canonical micro circuit, CMC) with neurovascular units (e.g., astrocytes), which regulate / alter changes in blood flow and deoxygenation (the haemodynamic response). The CMC model comprises four populations: 1) spiny stellate cells, 2) superficial pyramidal cells, 3) inhibitory interneurons, and 4) deep pyramidal cells. Each population is coupled to other populations via excitatory or inhibitory connections. In addition, a self-inhibitory connection – illustrated as a circular arrow – exists for each population (including excitatory populations) that models the gain or sensitivity of each population to its inputs, which is an inherent property of the dynamics of neuronal populations. EEG or MEG capture neuronal responses (modelled by interconnected CMC models), which may be distorted by the scalp – and are typically mixed with the activity of other nodes or sources. The fMRI signal derives from the haemodynamic part of the model. Changes in activity of neuronal populations (pre- or post-synaptic potentials) excite neurovascular coupling that in turn causes (e.g. by the release of nitric oxide) changes in blood flow. This is accompanied by changes in blood volume and a reduction in the level of deoxyhaemoglobin (dHb) in blood vessels, which give rise to the Blood Oxygen-Level Dependent (BOLD) response, measured using fMRI. Image credits: MRI scanner by Grant Fisher, TN from the Noun Project, CC BY 3.0.*

# Worked Example 1: Bayesian Model Reduction for structure learning

The aim of the first example is to showcase an application of BMR to modelling the neurovascular system using fMRI time series (55). This analysis used an fMRI dataset from a previously conducted experiment, investigating the neural response to attention to visual motion (56). Three brain regions (i.e., nodes) were identified that showed significant experimental effects (visual areas V1, V5 and the frontal eye fields, FEF) and representative timeseries were extracted from each region (i.e., feature selection). To explain the underlying generators of these timeseries, a DCM was specified comprising three canonical microcircuits (modelling within-region connectivity) that were coupled by between-region connections, creating a hierarchy of distributed processing nodes. The (presynaptic) input signal to each neuronal population in the CMC model comprises three components: two signals that are received from excitatory and inhibitory populations, and a third that is the summation of extrinsic inputs from distal regions. A weighted sum of these three signals (per neuronal population) form the input to the haemodynamic model. The weights that are inferred using DCM and BMR are then used to identify the optimal reduced coupling structure – the minimal set of neurovascular parameters – that best explain the data.

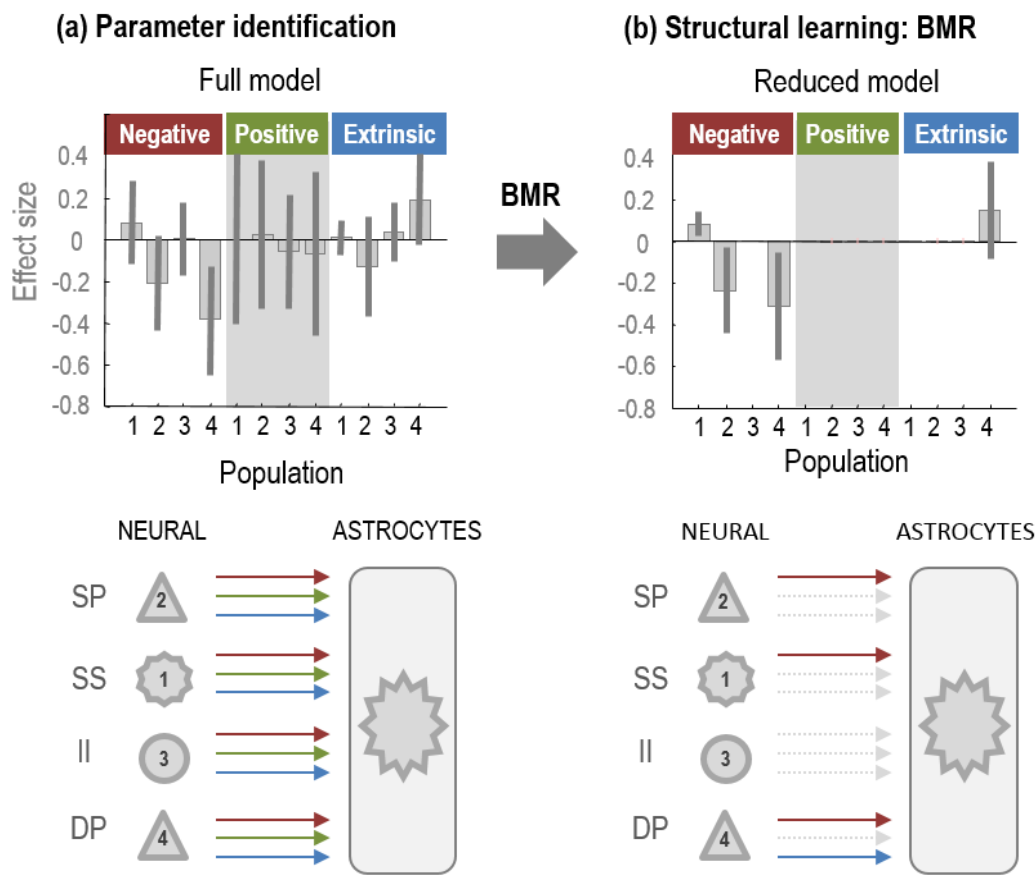

*Figure 3. **Structure learning of neurovascular coupling using BMR**. Panel (a) shows the posterior density over the neurovascular coupling parameters, modelled as the scaled sum of inhibitory (negative), excitatory (positive) and extrinsic signals to each population (i.e., SP: superficial pyramidal neuron, SS: spiny stellate cell, II: inhibitory interneuron and DP: deep pyramidal neuron). In panel (b), Bayesian model reduction (BMR) was performed using the posterior estimates of the parameters to identify a subset of parameters that best accounted for the data. The bar plots show the posterior expectations and 90% credible intervals of the neurovascular coupling parameters. Positive values indicate a positive contribution of a particular neuronal population to the overall vasodilatory signal, and negative values indicate a negative contribution of a neuronal population. The numbers associated with each population (1-4) correspond to the four neuronal populations in Figure 2. For this single subject's data, inhibitory inputs to the superior and deep pyramidal populations had a negative influence on the haemodynamic response. Smaller positive influences on the haemodynamic response were due to inhibitory inputs to spiny stellate cells and extrinsic inputs to deep pyramidal cells.*

The parameters of this model were estimated using the fMRI data (model identification), and those relating to neurovascular coupling are shown in Figure 3 (left). Next, BMR was applied, using an automatic greedy search over reduced models, to ask whether there was any sub-structure in the neurovascular model parameter space that could that could equally well account for the data, relative to the full model. The optimal reduced model is shown in Figure 3 (right), where inhibitory signals to three of the neuronal populations played a predominate role. From this single subject data, one could conclude that the origin of the BOLD signal can be primarily explained by the contribution of inhibitory inputs to superficial pyramidal cells, deep pyramidal cells and spiny stellate cells. In other words, the origin of the BOLD response is linked to the activity of inhibitory populations. This result was a technical illustration and confirming it would require a group study with a representative sample of the population. Nevertheless, it demonstrates that DCM and BMR enable the investigation of architectural questions about the underlying generators of fMRI time series using non-invasive recordings. In particular, the use of BMR with an automatic search allowed a fast and efficient search over a large model space.

# Worked Example 2: Bayesian fusion

This second study focussed on *Bayesian fusion* across neuroimaging modalities - MEG and fMRI. Data were collected using each modality, under the same cognitive task (an auditory oddball paradigm), to inform the neuronal and neurovascular / haemodynamic parts of the model, respectively (57). In the previous example, the neuronal part of the model had various parameters fixed (e.g., synaptic time constants), to estimate the neurovascular parameters using only fMRI data. In this second study, the objective was to develop a modelling scheme that utilizes the high temporal resolution of MEG to infer the parameters of the neuronal part of the model, before using the fMRI data to infer parameters of the neurovascular / haemodynamic part. Bayesian model selection was then used to infer the optimal reduced structure of the combined MEG-fMRI-informed model.

To enable flexible coupling of different neural and neurovascular models, this study introduced an interface between them called *neuronal drive functions* (Figure 4). These play a crucial role in the following analysis. First, active brain regions are identified from the fMRI data, using standard methods (i.e., statistical parametric mapping, SPM). The coordinates of activated regions are then used as spatial priors in the specification of the DCM for MEG model. By inverting this model, DCM for MEG identifies the generators of event related potentials (19) in terms of functional architectures; namely, condition-specific changes in intrinsic (within-region) and extrinsic (between-region) connectivity. Using the posterior expectations of the neuronal parameters, the canonical microcircuit DCM is then used to generate synaptic responses to each experimental condition (Figure 4b) – the neural drive functions. Finally, these functions (aligned to the timing of stimuli in the fMRI experiment) are used to drive a haemodynamic model of BOLD responses (Figure 4c) (57). The contribution of each population to the BOLD response is parameterised ($\beta_n$), where these parameters are estimated from the fMRI data. By specifying models with different parametrisations, this procedure enables comparison of different models of neurovascular coupling (BMS).

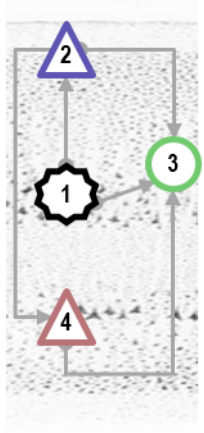
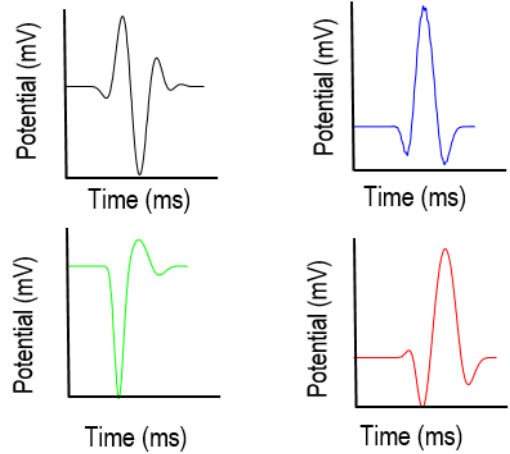
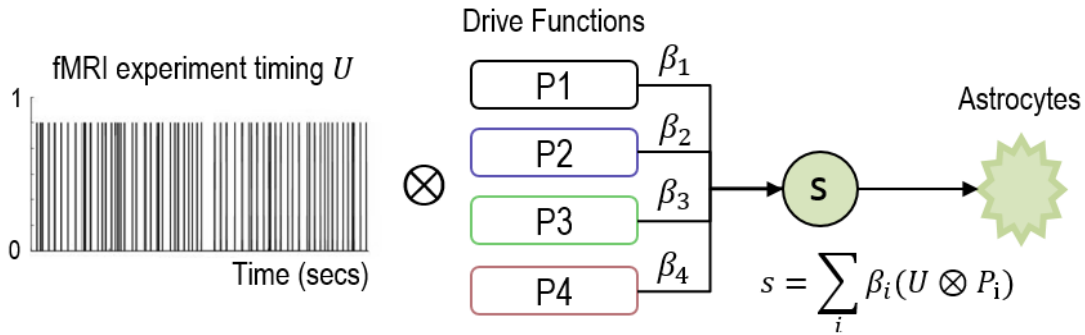

*Figure 4. Multi-modal DCM for MEG-fMRI pipeline. a) The neural network is modelled with a canonical microcircuit (CMC) for each brain region (i.e., node). The parameters of the CMC model are estimated using DCM for MEG, with spatial priors determined by a mass-univariate regression analysis of the fMRI data. b) Condition specific neuronal response functions are simulated from the CMC model using the estimated parameters, associated with each population (P1,...,P4) and each experimental condition. The top row shows (impulse) response functions from neuronal populations SS and SP, and the bottom row shows (impulse) response functions from populations II and DP, for a single experimental condition. c) The inputs to the haemodynamic model are calculated by summing and temporally shifting (convolving) the neuronal response functions with the timeline of the experimental stimulation in the fMRI paradigm. The ensuing neuronal drives from each population are scaled by parameters β to form a neurovascular signal, s(t), which forms the input to the haemodynamic model. The haemodynamic model is fitted to the fMRI time series and hypotheses are tested by comparing the evidence for reduced models with different combinations of β parameters switched on or off.*

This study also provides an opportunity to illustrate the use of pre-defined model spaces. Whereas in the first example, an automatic search was conducted over reduced models; here, a set of carefully specified models were compared to test specific experimental questions or factors. The ensuing set of models – the model space – characterised neurovascular coupling in terms of four factors or questions. In brief, the factors were: (i) whether there were presynaptic (37) or postsynaptic (54) contributions to the neurovascular signal, (ii) whether neuronal responses from distal regions excite neurovascular coupling (38), (iii) whether the neurovascular coupling function was the same across all regions or should be defined in a region-specific way (39), and finally (iv) whether a first or second order differential equation should be used for the dynamics of the neurovascular system, to determine if there was any delay between neuronal activity and the ensuing BOLD response (40,41). A total of 16 candidate models, with different combinations of these four factors, were evaluated using Bayesian model comparison. For each of the four experimental questions, the 16 models were grouped into families (e.g. all models with presynaptic vs postsynaptic input) and the probability for each family computed.

In this illustrative single subject study, the results of family wise BMS on each group of models identified strong evidence (with nearly 100% posterior confidence) for the following:

(i) The BOLD effect was caused by presynaptic signals, which is in line with the findings of (37), which found mean neuronal firing rates (presynaptic signals) induced BOLD signals.

(ii) Regional neuronal drives to haemodynamic responses induced vascular perfusion. This is consistent with general opinion that local neuronal activity induces BOLD contrast.

(iii) The strength of neurovascular coupling was region specific. This is in agreement with invasive recordings using animal studies suggesting that neurovascular coupling varies from brain region to region (39).

(iv) The response of the BOLD response to neuronal activity was instantaneous, rather than delayed.

This example illustrates some of the intricacies of structure learning using DCM and BMS. First, hypotheses about biophysical processes can be expressed formally as models, and in particular as a factorial design (i.e., a model space that can be partitioned among different factors or attributes). Second, data from different neuroimaging modalities can be combined using Bayesian fusion, where different parts of the model are informed by different modalities. For flexibility and efficiency of model inversion, neuronal drive functions were introduced to act as a bridge between neural and haemodynamic models. Finally, each experimental question can be addressed through family-wise Bayesian model comparison. In summary, these two example applications illustrate the application of DCM, BMR and BMS to structure learning based on multi-modal neuroimaging data.

# Discussion

In this paper, we have reviewed a suite of recently developed methods for structure learning in the context of coupled dynamical systems. In particular, we showcased applications in neuroscience using *dynamic causal modelling (DCM)* – the Bayesian inversion of biophysically informed coupled dynamical systems, and *Bayesian model selection (BMS)* and *reduction (BMR)* for assessing the evidence for different models or hypotheses. To date, these tools have mainly been applied in the context of cognitive and clinical neuroscience – to unravel the functional architectures underlying neuroimaging data. DCM offers an efficient way to estimate the parameters of large-scale dynamical system based on a gradient descent of a variational free energy functional. This functional inherently scores different candidate architectures and coupling functions in terms of a trade-off between accuracy and model complexity.

In a general setting, DCM, BMS and BMR offer efficient pipelines for modellers to identify coupled dynamic systems in an evidence-based fashion. DCM has been applied to a wide range of problems including parameter estimation for deterministic (58) and stochastic (36,59) dynamical systems using time, frequency or time-frequency domain information. In addition, DCM has been found to be useful for studying large networks, based on the centre manifold theorem (44) and parameter estimation of dynamical systems on a manifold (60). These examples demonstrate that structure learning based on DCM, BMR, and BMS provide a general and efficient method that can be applied to a wide range of modelling problems of real word physical systems.

One might ask whether dynamic causal modelling has real word clinical applications, beyond research. For instance, would it be possible to use DCM as a part of biological control system, to suppress or prevent unwanted activity in a diseased brain (e.g., epileptic seizures, or the symptoms of Parkinson's disease)? This question has been addressed in the setting of Parkinson's disease (61); however, there is a long road ahead before these models are sufficiently well developed and validated to be used in the treatment of neurological disorders. In a research context, an avenue receiving much attention is how to validate theories of brain function based on predictive coding and more generally the Bayesian brain hypothesis – and in particular, how to identify the mechanisms of information transfer between layers of cortex (24,25,62). Here, tools such as the CMC model, variational Laplace and structure learning using BMS and BMR are likely to prove useful.

**Acknowledgments**
None

**Funding Statement**
The Wellcome Centre for Human Neuroimaging is supported by core funding from Wellcome [203147/Z/16/Z].

**Data Accessibility**
Code demonstrating the methods described in this paper can be found in the freely available SPM software package – https://www.fil.ion.ucl.ac.uk/spm/. After installation, type DEM (short for Demo) and press enter, to view all available demos.

**Competing Interests**
None

**Authors' Contributions**
AJ wrote the manuscript and prepared the figures. PZ, VL and KJF contributed to planning and editing the manuscript.

# Table

**Table 1:** glossary of terms that are used in the paper and their description.

| Term | General Description |
|---|---|
| Bayesian belief updating | Bayesian belief updating is the updating of probabilities as new data are acquired. An example is found in the context of inverting nonstationary EEG time series that undergoes transition into and out of paroxysmal activities (15). In this case, EEG data is first divided into several locally (quasi) stationary segments. Then, DCM is performed on the first segment to infer posterior parameter estimates. The ensuing estimate becomes the prior for the subsequent segment and so on. This is Bayesian belief updating over time. A trajectory of parameters over segments can then be constructed to characterise the dynamics of the nonstationary data. |
| Bayesian fusion and multimodal data integration | Bayesian fusion makes combined inferences about a (physical) system based on different kinds of observation, in order to model the system's dynamics. For instance, Bayesian fusion of models fitted to fMRI and M/EEG data can be used to understand the function of neurovascular coupling. The spatial specificity of fMRI is employed to localise active neuronal sources. This spatial information is used as a prior for source localisation within a model of M/EEG data, which is employed to estimate neuronal |
| | parameters. The estimated neuronal parameters are taken as priors for the inversion of haemodynamic responses (i.e., Bayesian fusion) from fMRI time series. Therefore, combining these modalities provides a better understanding of neurovascular coupling than could be derived from either modality independently. |
| Bayesian model comparison | Quantification of the relative evidence for different models of the same data. Typically this is expressed as a Bayes factor, which is the ratio of the model evidence (marginal likelihood) for each model relative to a selected comparison model. |
| Bayesian model reduction (BMR) | A statistical method for rapidly computing the posterior probability density over parameters and the log model evidence for a reduced model, given the parameters and log model evidence of a full model. Here, reduced and full models differ only in their priors. Under Gaussian assumptions, this has an analytic form. |
| Bayesian model selection (BMS) | The selection of one or more of models with the highest evidence from a set of candidate models following Bayesian model comparison. |
| Canonical micro circuit (CMC) | The CMC is a biologically informed microcircuit model of laminar connectivity in the cortical column (17,34). It comprises four populations whose activity can replicate a realistic pattern of M/EEG signals. Each population generates postsynaptic potentials (modelled by second order differential equations) induced by presynaptic firing rates from external sources (interregional or distal populations, and/or exogenous inputs). These postsynaptic potentials generate presynaptic firing rates (via a sigmoid transformation), which in turn excite or inhibit other populations. |
| Compressive sensing based dynamical system identification (9) | Compressive sensing is a well-established signal processing method for reconstructing sparse signals from data. Recently it has been applied for identification of fully observed dynamical systems (i.e., all states of the system are measured) with the assumption that the evolution of states can be modelled using (linear in parameters) power series and that the underling structure (coupling between nodes) of the system has a sparse form. Given these assumptions, it can be shown that the relation between measured timeseries and model states can be formulated as a linear function, parametrised by unknown parameters. Compressive sensing is then applied to find the sparse parameter vector that best explains the data. |
| Cortical column | The human cortex can be approximately divided into cylinders of diameter 500 $\mu m$ where neurons within each cylindrical column activate in response to a particular stimulus. |
| Data assimilation (63) & model identification (64) | Data assimilation is a term that was coined by meteorologists in the mid-20$^{th}$ century. It uses (and combines) a wide range of mathematical methods such as nonparametric statistical models (auto regressive models), nonlinear Kalman filters, statistical interpolation, nonlinear time series analysis and nonlinear system identification to establish models useful for weather prediction (65). Model identification is a more general term that accounts for constructing models describing a phenomenon, which includes model construction and parameter estimation. |
| Data feature | In general, quantities that are useful for distinguishing/discriminating different regimes of a system are referred to as data features. In DCM, conventionally, models are either fitted to the raw data directly, or to data features such as autoregressive coefficients that model the power spectrum, power spectral densities (from time-frequency analyses) and principal components/modes and the phase of data. Extracting particular features of the data for modelling is known as feature selection, and is typically conducted as an initial step in the modelling pipeline. |
| Effective connectivity | Effective connectivity can be understood in relation to functional and structural connectivity (66). Functional connectivity is defined as a statistical dependency (such as correlation or transfer entropy) between multi-channel brain data (e.g., fMRI, M/EEG). Structural connectivity refers to the strength of anatomical connections between brain regions and can be estimated, for instance, using diffusion tensor imaging. Effective connectivity is the directed effect of neural populations on each other, under a particular experimental setting or task. Inferring effective connectivity typically requires combining brain imaging data with a biologically informed model of brain activity. This enables one to elucidate (i) underlying biological generators of |
| | the data (ii) how different experimental inputs (or conditions) alter the effective connectivity. |
| Electroencephalography (EEG) and Magnetoencephalography (MEG) | M/EEG are non-invasive neuroimaging techniques that capture dynamics of neuronal activity with millisecond temporal resolution (on the same order as the temporal dynamics of synaptic activity). EEG captures ensemble neuronal membrane potential voltages using grids of electrodes placed on the scalp, whereas MEG measures accompanying fluctuations in magnetic fields that can be captured using arrays of magnetometers (known as superconducting quantum interference devices). The MEG signal is subject to less distortion by the skull and scalp than EEG. |
| Functional magnetic resonance imaging (fMRI) | fMRI is a non-invasive neuroimaging technique that measures changes of blood flow and oxygenation with a fine spatial resolution (effectively up to $0.5\ mm$) due to neuronal activation and the neurons' subsequent consumption of oxygen. fMRI measures the blood-oxygen-level dependent (BOLD) response to brain activity. Changes in the measurement at each location (voxel) forms a time series, which is analysed using the analytic techniques reviewed here. |
| Haemodynamic response | This describes the process by which neuronal activation causes changes in blood flow, blood vessel volume and the dynamics of deoxyhaemoglobin. It can be modelled by a dynamical system known as the extended Balloon model (33). |
| Ill-posed problem and model identification in DCM | The identification of biological mechanisms generating measurable brain data is, in general, an ill posed problem. This is predominately because (i) for most brain imaging techniques only partial, indirect, noisy and nonlinearly mixed data are available (e.g., EEG is mostly generated by the activity of a small proportion of pyramidal populations and has poor spatial resolution, fMRI is an indirect measure of neuronal activity with poor temporal resolution due to the smoothing effect of haemodynamics) and (ii) given the complexity of biological models, there are typically many parameters giving rise to similar data. With DCM, the ill-posed problem is addressed by setting suitable prior constraints on model parameters. |
| Inhibitory, excitatory and pyramidal neuronal population (31) | Intuitively, postsynaptic potentials generated by inhibitory interneurons reduce the depolarisation and subsequent activity of target populations. Conversely, excitatory interneurons increase the activity of target populations. Pyramidal cells are excitatory and can be found in superficial and deep layers of the cortical column. |
| Kullback-Leibler (KL) divergence | This is a measure of the difference between two distributions. However, it is not a metric since it is not commutative (62). As explained in this paper, it can be used to score the complexity of a model (the difference between posterior and prior probability densities over model parameters). |
| Log likelihood | The log probability of the observed data given the model, under a particular set of parameters (25,30). In this paper, it serves as the accuracy of a model in terms of the probability of producing the observed data features. Integrating over the unknown model parameters gives the model evidence (or marginal likelihood). |
| Neurovascular coupling and neural drive function | Neurovascular coupling refers to physiological pathways that enable communication between neurons and blood vessels (55). A neuronal drive function is the scaled sum of neuronal activity, which is estimated using DCM for electrophysiological (e.g., EEG/MEG) recordings, and forms the input to a model of the neurovascular system (56). |
| Model architecture | Model architecture – in this paper – refers to a dynamical system distributed on a graph, consisting of nodes (e.g., neuronal populations) and edges; i.e., coupling or connections between the nodes. The dynamics of each node are governed by differential equations. In DCM, connections are set as being present (informed by the data) or absent (fixed at zero) by specifying the variance of Gaussian prior probability densities. |
| Statistical Parametric Mapping (SPM) (7) | SPM is freely available software for analysing brain imaging data such as fMRI, MEG and EEG. It includes statistical routines (e.g., general linear model, random field theory, variational Bayes, voxel based morphometry, statistical hypothesis testing, statistical signal processing, to name but a few). SPM also refers to a method for producing maps of parametric statistics, to test for distributed neural responses over the brain. The SPM software package also includes the dynamic causal modelling |
| | (DCM) toolbox, which enables the modelling of the underlying biological generators of neuroimaging data. |
| Time-evolving dynamical Bayesian inference (10) | A form of sequential Bayesian inference, developed to infer time-dependent coupling of noise-driven weakly coupled dynamical systems (e.g., coupled limit-cycle, phase, and chaotic oscillators). Given successive segments (in time) of data, the algorithm maximises the likelihood (and posterior probability density over unknown parameters) in the first segment. Then, a corrected form of posterior estimate of the parameters is considered as the prior of parameters in the next segment and so on. With this method, belief updating from one segment to the next is accompanied by applying a diagonal (correction) matrix to the covariance estimate of the parameters, preventing correlations among parameters propagating over time. |
| Tracking based parameter identification in dynamical system (8) | Constant parameters in a dynamical system have zero temporal evolution. Therefore, they can be considered as state variables with trivial dynamics. The trivial dynamics of parameters can be absorbed into a state space model by augmenting the original equations of the system (thereby named an augmented dynamical system). A filtering method (e.g., nonlinear Kalman filter) can then be applied to reconstruct (estimate) the dynamics the augmented system that includes slowly fluctuating parameters. The mean and variance of the resulting parameter estimates can be taken as posterior estimates of constant parameters of the original dynamical system. |
| Variational Laplace | Variational Bayes under the Laplace assumption. A scheme for approximate Bayesian inference in which the priors and approximate posteriors are (multivariate) normal. In contrast to most variational schemes, variational Laplace uses a gradient ascent on variational free energy. This is important because it eschews the need for analytic solutions and the use of conjugate priors. This means that model inversion uses exactly the same optimisation scheme for any given generative model. |

# Supplementary material

*Jafarian A, Zeidman P, Litvak V and Friston K, Structure Learning in Coupled Dynamical Systems and Dynamic Causal Modelling*

This brief note explains the parameter estimation routine in DCM. Detailed explanation can be found in (1,2). A generic form of a nonlinear model with unknown parameter vector $\theta$ is as follows:

$$y = g(\theta) + e \tag{A1}$$

In equation (A1), $e$ is an additive error term with Gaussian distribution $e \sim N(0, C_y)$ (where $C_y$ is a covariance matrix). The inverse of the error covariance matrix is called the precision matrix and is denoted by $\Pi_y$. The precision matrix can be decomposed using known precision basis functions $Q_i$ as follows:

$$\Pi_y = C_y^{-1} = \sum_i \exp(\lambda_i) Q_i \tag{A2}$$

In equation (A2), scalar $\lambda_i$ is called a hyperparameter (since it is used to define the distribution of the error term) and $i$ is the number of the precision component. The aim of model *inversion* is to infer the posterior probability of parameters $\theta$ and hyperparameters $\lambda$ given their (normal) prior densities which are denoted by $p(\theta) = N(\theta; \mu_\theta, C_\theta)$ and $p(\lambda) = N(\lambda; \mu_\lambda, C_\lambda)$, respectively. An iterative optimisation scheme is used in DCM that searches for settings of the parameters that maximise the log model evidence $\ln p(y)$ (1,2). However, since in practice the log model evidence cannot be computed exactly, and an approximation called the negative variational free energy functional $F$ (or evidence lower bound) is used and is returned by DCM. The negative variational free energy, hereafter the free energy, is used as the basis for comparing the evidence for different candidate models (Bayesian model selection).

The scheme for estimation of model parameters in DCM is called variational Bayes method under Laplace approximation (a.k.a, variational Laplace), and is employed to approximate posterior densities of parameters with Gaussian distributions. The approximation can be evaluated in terms of the difference ($KL$ divergence) between the true posterior $p(\theta, \lambda | y)$ and the approximated Gaussian posterior $q(\theta, \lambda)$ over the parameters. However, since the true posterior is not known, the difference cannot be computed directly, and the goodness of the approximation needs to be evaluated indirectly through optimisation of the free energy. This follows because the log model evidence can be expressed as the algebraic sum of the negative free energy $F$ and the KL divergence, given in Equation 2 of the main text and re-expressed here for convenience:

$$\ln p(y) = F + KL[q(\theta, \lambda) || p(\theta, \lambda | y)] \tag{A3}$$

As the left hand side of this equation - the log model evidence $\log p(y)$ – is a fixed value, maximising the free energy $F$ will minimize the difference between the approximate and true parameter densities.

Estimation of the densities in DCM rests upon a mean field approximation, meaning that probability densities over the parameters and hyperparameters are treated as independent and can therefore be factorised. The posterior density over the parameters $q(\theta, \lambda)$ factorise as follows:

$$q(\theta, \lambda) = q(\theta)q(\lambda)$$
$$q(\theta) = N(\theta, m_\theta, S_\theta) \quad \text{(A4)}$$
$$q(\lambda) = N(\lambda, m_\lambda, S_\lambda)$$

This allows estimation of parameters and hyperparameters to be carried out separately. Under the mean field approximation, the free energy functional can be written as follows:

$$F = \int \int q(\theta)q(\lambda) \ln p(y, \theta, \lambda) d\theta d\lambda - \int q(\theta) \ln q(\theta) d\theta - \int q(\lambda) \ln q(\lambda) d\lambda \quad \text{(A5)}$$

In DCM, to calculate the free energy in equation (A5), the function $\ln p(y, \theta, \lambda)$ is approximated around the estimated posterior mean of unknown parameters using a second order Taylor series expansion - formally known as a Laplace approximation (1). In effect, equation (A5) can be expressed in a closed form (i.e., analytically) in terms of Gaussian functions.

To find the posterior density over the parameters, iterative maximization of the Laplace approximation of the following energy integrals is performed by the DCM software:

$$I_\theta = \int L(\theta, \lambda) q(\lambda) d\lambda$$
$$I_\lambda = \int L(\theta, \lambda) q(\theta) d\theta \quad \text{(A6)}$$

In equation (A6), $L(\theta, \lambda) = \log[p(y|\theta, \lambda)p(\theta)p(\lambda)]$ is the likelihood function. During the ensuing iterations, the posterior mean of the parameters (in the same way for hyperparameters) are updated using the following rule:

$$\mu_\theta^{new} = \mu_\theta^{old} + \delta\mu_\theta \quad \text{(A7)}$$

In the equation (A7), $\delta\mu_\theta$ is the step size for updates to the model parameters. This is defined as $\delta\mu_\theta = -H_\theta^{-1} J_\theta$ where $J$ and $H$ are the gradient ($J_\theta = \frac{\delta I_\theta}{\delta\mu_\theta}$) and Hessian ($H_\theta = \frac{\delta^2 I_\theta}{\delta\mu_{\theta_i}\delta\mu_{\theta_j}}$) of the free energy functional. In effect, equation (A7) resembles a Newton update method. Intuitively, in the region that the gradient changes slowly, changes to the parameters are large to avoid an unnecessary search, making the proposed algorithm efficient. The posterior covariance of the parameters is defined as negative inverse curvature $S_\theta = -H_\theta$ (for hyperparameters $S_\lambda = -H_\lambda$). The estimation scheme iterates over the parameters and hyperparameters until the free energy quantity ceases to change significantly.